\documentclass[10pt,a4paper]{article}
\usepackage[utf8]{inputenc}
\usepackage[english]{babel}
\usepackage{amsmath}
\usepackage{amsfonts}
\usepackage{amssymb}

\RequirePackage[colorlinks,citecolor=blue,urlcolor=blue,linkcolor=blue]{hyperref}

\usepackage[top=2cm,bottom=2cm]{geometry}
\usepackage{authblk}

\usepackage{feynmp}
\newcommand{\pd}{\partial}
\newcommand{\ce}{\varepsilon}

\begin{document}

\title{On anomalies in effective models with nonlinear symmetry realization}

\author[1,2]{\href{https://orcid.org/0000-0001-9326-6905}{Andrej Arbuzov}\footnote{\href{mailto:arbuzov@theor.jinr.ru}{arbuzov@theor.jinr.ru}}}

\author[1,2]{\href{https://orcid.org/0000-0001-7099-0861}{Boris Latosh}\footnote{\href{mailto:latosh@theor.jinr.ru}{latosh@theor.jinr.ru}}}

\affil[1]{Bogoliubov Laboratory of Theoretical Physics, JINR, Dubna 141980, Russia}
\affil[2]{Dubna State University, Universitetskaya str. 19, Dubna 141982, Russia}

\maketitle

\begin{abstract}
  Anomalous features of models with nonlinear symmetry realization are addressed. It is shown that such models can have anomalous amplitudes breaking of its original symmetry realization. An illustrative example of a simple models with a nonlinear conformal symmetry realization is given. It is argued that the effective action obtained via nonlinear symmetry realization should be used to obtain an anomaly-induced action which is to drive the low energy dynamics.

\end{abstract}

\section{Introduction}

Conformal symmetry plays a special role in contemporary physics. Firstly, the Standard Model 
of particle physics is almost conformally invariant, as it contains a single primary dimensionful 
parameter, namely, the Higgs boson mass. Secondly, the conformal group is related to the spacetime 
symmetry via the Ogievetsky theorem~\cite{Ogievetsky:1973ik}. The theorem states that the 
infinite-dimensional coordinate transformation algebra is generated via commutation of generators 
from the Lorentz and conformal groups. At the same time, the conformal group cannot be directly 
realized, as the corresponding Noether currents do not exist in Nature. Consequently, at the coordinate 
level the conformal symmetry can appear through a non-linear realization~\cite{Borisov:1974bn}. 
At the level of physical fields the conformal group can also be realized 
in a non-linear way~\cite{Arbuzov:2017sfw,Arbuzov:2010fz,Isham:1970xz,Isham:1970gz,Isham:1971dv}.

This paper addresses, perhaps, the simplest model with a non-linear realization of the conformal 
symmetry at the level of physical fields~\cite{Arbuzov:2017sfw}. 
The model was presented in paper~\cite{Arbuzov:2019rcl}, 
it is given by the following effective action:
\begin{align}\label{the_effective_action}
  \Gamma = \int d^4 x \Bigg[ &\cfrac12 \left\{1+\cfrac{\sigma^{(\alpha)} \sigma^{(\alpha)}}{\ce^2} ~E_2^2\left(\cfrac{\psi}{\ce}\right)\right\} \pd_\mu\psi\, \pd^\mu\psi \\
    & +\cfrac12~E_1^2\left(\cfrac{\psi}{\ce}\right)~ \pd_\mu\sigma^{(\alpha)} \pd^\mu\sigma^{(\alpha)} -\cfrac{1}{\ce}~ E_1\left(\cfrac{\psi}{\ce}\right)E_2\left(\cfrac{\psi}{\ce}\right) ~\pd_\mu\psi~ \sigma^{(\alpha)}\pd^\mu \sigma^{(\alpha)} \Bigg]. \nonumber
\end{align}
The action is recovered via symmetry 
principles~\cite{Volkov:1973vd,Ivanov:2016lha,Coleman:1969sm,Callan:1969sn,Weinberg:1968de} and 
describes Goldstone scalar modes $\psi$ and $\sigma^{(\alpha)}$, $(\alpha)=0,1,2,3$ which arise 
due to a spontaneous breaking of the conformal symmetry down to the Lorentz group. Here index $(\alpha)$ is not a Lorentz one, but it is associated with conformal group generators. 
In this expression $\ce$ is the symmetry breaking energy scale. Functions $E_1$ and $E_2$ are 
defined as
\begin{align}
  E_1(x)&\overset{\text{def}}{=} \cfrac{e^x-1}{x}\, , & E_2(x)&\overset{\text{def}}{=} \cfrac{e^x-x-1}{x^2} \, . 
\end{align}
Finally, it should be noted that within such a simple model $\psi$ may be treated as a dilaton associated with the spontaneously broken conformal symmetry. 

The main aim of this paper is to study anomalous behavior of this model. Below it is shown that 
the model admits anomalous amplitudes which do not respect the non-linear conformal symmetry
realization. Consequently, it is argued that the effective action~\eqref{the_effective_action} 
should be substituted by the anomaly-induced effective action which, in turn, drives the low-energy
dynamic of the theory. 

Anomalies are crucial for effective theories. Perhaps, the most well-known example of anomalous 
behavior is given by the axial anomaly~\cite{Adler:1969gk,Adler:2004qt}. Namely, the following 
diagram reveals anomalous properties:
\begin{align}
  \nonumber \\
  \begin{gathered}
    \begin{fmffile}{pic01}
      \begin{fmfgraph*}(50,50)
        \fmfbottom{B1,B2}
        \fmftop{T}
        \fmf{fermion,tension=.3}{L,R,U,L}
        \fmf{photon}{B1,L}
        \fmf{photon}{B2,R}
        \fmf{photon}{U,T}
        \fmflabel{$\gamma^\sigma\gamma^5$}{T}
        \fmflabel{$\gamma^\mu$}{B1}
        \fmflabel{$\gamma^\nu$}{B2}
      \end{fmfgraph*}
    \end{fmffile}
  \end{gathered}
\end{align}
The divergent part of this diagram cannot be renormalized, but it is canceled out completely when 
a symmetric diagram with $\mu\leftrightarrow\nu$ is taken into account. On the contrary, the finite
parts of such diagrams do not vanish and generate a new physical interaction corresponding to a real
effect. 

Anomalies in a fundamental theory lead to undesirable effects such as unitarity 
violation~\cite{Gross:1972pv}, so it is safe to assume that fundamental theories are 
free from them. For instance, within the Standard Model all axial anomalies are canceled 
out~\cite{Bouchiat:1972iq,Minahan:1989vd}.

The same logic cannot be applied for effective models.
First and foremost, the notion of an anomaly is traditionally used within renormalizable field theories. Within non-renormalizable models new interactions are generated by loop corrections, so the notion of an anomaly should be refined.

In this paper we discuss anomalies in the context of effective models. Because of this it is safe to assume that such models, despite being non-renormalizable, have suitable underlying fundamental theories. The nonlinear symmetry realization allows one to account for a non-trivial dynamics existing within the fundamental theory. Therefore all divergences appearing within an effective theory are expected to be regularized with the fundamental theory.

In that context it is reasonable to study operators generated at the loop level within effective models. Some of these operators do not respect the original nonlinear realization of the symmetry and they should be understand as footprints of further radiative symmetry breaking occurring in the fundamental model. Thus, in the context of effective models we use the notion of an anomaly to refer to the presence of such operators. Let us highlight one more time, that the existence of such operators do not introduces new pathologies. Effective models themselves have a limited applicability domain, so they are free from the corresponding pathologies, while a fundamental theory can always be safely assumed to be free from pathologies.

The best example of an effective model anomaly is given by the $\sigma$-model for pions~\cite{Bell:1969ts}. The model 
admits anomalous diagrams which are essential for the observed $\pi^0\to\gamma\gamma$ 
decay~\cite{Adler:1969gk,Bell:1969ts}. In full analogy with the model to be addressed, 
the pion $\sigma$-model is recovered via nonlinear realization of the chiral 
symmetry~\cite{Weinberg:1968de}. Its anomalous amplitudes do not respect the non-linear 
symmetry realization and thereby violate the original symmetry of the effective action.

Such considerations involve three features of anomalies. Firstly, anomalous contributions describe 
real physical effects which can be probed empirically. Secondly, anomalies break the original model
symmetry that exists in the effective action at the classical level. Consequently, the effective action
recovered via symmetry principles should be used to obtain an anomaly-induced effective action which, 
in turn, drives the low energy phenomenology. Most importantly, the anomaly-induced effective action 
may not respect the original model symmetry. Finally, these conclusions can be universally applied 
to all models with non-linear symmetry realizations including the model addressed in this paper.

The rest of the paper is organized as follows. Firstly, we show that the one-loop three-point 
function generated by the effective action~\eqref{the_effective_action} at the first order 
in $\ce^{-1}$ is anomalous off-shell and vanishes on-shell. Secondly, we show that the one-loop
four-point function generated by the effective action at the first order in $\ce^{-1}$ is anomalous
on-shell, but these divergences can be eliminated completely via Ward identities. Finally, we show 
that a similar argument does not hold for one-loop four-point functions evaluated at the second order 
in $\ce^{-1}$ by the effective action. Consequently, we are going to show that the model admits
anomalous features and to recover its anomaly-induced effective action up to $\ce^{-4}$ order. 
The paper concludes with a discussion of results and their role within gravity models with non-linear 
symmetry realization.

\section{Anomalous contributions}

A more detailed comment on the studied model \eqref{the_effective_action} is due. As it was noted above, the model is recovered via symmetry reasoning. Models with nonlinearly realized symmetries, including nonlinear sigma models, were actively studied in a large number of papers. An explicit method generating a Lagrangian with a given nonlinear realization was found in a series of papers \cite{Weinberg:1968de,Coleman:1969sm,Callan:1969sn}.

The method is implemented as follows. One starts with a big ground $\mathcal{G}$ and its subgroup $\mathcal{H}$. As it is well shown in \cite{Weinberg:1968de,Coleman:1969sm,Callan:1969sn} there is an explicit way to construct an action of the big group $\mathcal{G}$ both on the small group $\mathcal{H}$ and on a corresponding quotient $\mathcal{G}/\mathcal{H}$. This generates two realizations of $\mathcal{G}$; first one represents $\mathcal{G}$ on the small group $\mathcal{H}$, while the second one defines a representation of $\mathcal{G}$ on $\mathcal{G}/\mathcal{H}$. The master formula for such nonlinear realizations is given in multiple papers \cite{Weinberg:1968de,Coleman:1969sm,Callan:1969sn,Volkov:1973vd,Ivanov:2016lha,Arbuzov:2019rcl}, so it will not be presented here for the sake of briefness.

Most importantly, in papers \cite{Coleman:1969sm,Callan:1969sn} it was explicitly shown that there is a way to recover covariant derivatives of fields subjected to such a nonlinear realization. One can recover their explicit form knowing only on the big group $\mathcal{G}$ and the small group $\mathcal{H}$. Therefore an explicit form of the Lagrangian in a model with a given nonlinear symmetry realization can be obtained without a reference to an explicit form of the transformation laws (see our previous paper \cite{Arbuzov:2019rcl} with a more detailed discussion).

In paper \cite{Arbuzov:2019rcl} this algorithm was implemented for the conformal group. We have shown that the model \eqref{the_effective_action} is generated via a nonlinear symmetry realization on subgroup generated by conformal distortions. To be exact, one has to construct a quotient of the conformal group $\mathcal{C}(1,3)$ on a subgroup generate by conformal distortion generators $\mathcal{D}$, $\mathcal{K}_{(\mu)}$. Here $\mathcal{D}$ is a uniform conformal distortion operator, $\mathcal{K}_{(\mu)}$ are operators corresponding to conformal distortions orthogonal to the corresponding coordinate lines, and $(\mu)=0 \cdots 3$ is the index which numbers these operators. The index is put into brackets, as it must not be confused with the standard Lorentz indices. The paper \cite{Arbuzov:2019rcl} is completely devoted to a detailed discussion of an implementation of nonlinear symmetry realization to gravity. A more detailed discussion of the model \eqref{the_effective_action} lies beyond the scope of this paper and can be found elsewhere \cite{Arbuzov:2019rcl}.

Summing up what has been said, the model \eqref{the_effective_action} is constructed in a way to respect a certain nonlinear realization on the conformal group \cite{Arbuzov:2019rcl}. The explicit way to generate it together with a more detailed discussion of the corresponding nonlinear transformations can be found in \cite{Arbuzov:2019rcl}. Still, we present an explicit form of the transformation law below \eqref{the_transformation_law_explicitly}, as it is relevant in the context of the Ward identities.

To proceed with the aims we cast the effective action \eqref{the_effective_action} in terms of $\ce^{-1}$ expansion:
\begin{align}\label{the_effective_action_momentum_expansion}
  \begin{split}
    \Gamma = \int d^4 x \Bigg[ & \cfrac12~(\pd\psi)^2 + \cfrac12~(\pd\sigma)^2 + \cfrac{1}{\ce}\,\psi\,\left[ \pd_\mu\sigma^{(\alpha)}\pd^\mu\sigma^{(\alpha)} +\cfrac12\,\sigma^{(\alpha)} \square\sigma^{(\alpha)} \right]  \\
      & +\cfrac{1}{4\ce^2}\, \left\{\cfrac12\, \sigma^2 (\pd\psi)^2+\cfrac{7}{6} \,\psi^2(\pd\sigma)^2-\cfrac53\, \psi\pd_\mu\psi\cdot\sigma\pd^\mu\sigma\right\} + O(\ce^{-3}) \Bigg].
  \end{split}
\end{align}
It should be noted that the expansion in terms of the inverse symmetry breaking energy $\ce^{-1}$ 
matches the expansion in $N$-particle interaction functions. In other words, the term describing 
$N$-particle interaction belongs to $\ce^{2-N}$-order of the expansion. Because of this the effective 
action \eqref{the_effective_action} is not a momentum expansion.

Up to the order $\ce^{-2}$ tree-level diagrams of the effective action are
\begin{align}
  \begin{gathered}
    \begin{fmffile}{pic02}
      \begin{fmfgraph*}(20,20)
        \fmfleft{L}
        \fmfright{R}
        \fmf{dots}{L,R}
        \fmflabel{$\psi$}{L}
        \fmflabel{$\psi$}{R}
      \end{fmfgraph*}
    \end{fmffile}
  \end{gathered}
  &\hspace{.5cm}=\cfrac{i}{k^2} ~, &
  \begin{gathered}
    \begin{fmffile}{pic03}
      \begin{fmfgraph*}(20,20)
        \fmfleft{L}
        \fmfright{R}
        \fmf{dashes}{L,R}
        \fmflabel{$\sigma^{(\alpha)}$}{L}
        \fmflabel{$\sigma^{(\beta)}$}{R}
      \end{fmfgraph*}
    \end{fmffile}
  \end{gathered}
  &\hspace{.8cm} = \cfrac{i}{k^2} ~ \delta^{(\alpha)(\beta)} ~,
\end{align}
\begin{align}
  \begin{gathered}
    \begin{fmffile}{pic04}
      \begin{fmfgraph*}(45,45)
        \fmfset{arrow_len}{.3cm}
        \fmfleft{L}
        \fmfright{R1,d,R2}
        \fmf{dots_arrow,label=$k$,label.side=left}{L,V}
        \fmf{dashes_arrow}{R1,V}
        \fmf{dashes_arrow}{R2,V}
        \fmflabel{$p,(\alpha)$}{R1}
        \fmflabel{$q,(\beta)$}{R2}
        \fmfdot{V}
      \end{fmfgraph*}
    \end{fmffile}
  \end{gathered}
  &=-\cfrac{i}{2\ce}\left[p\cdot q + \cfrac12~k^2\right]~\delta^{(\alpha)(\beta)}~, \nonumber\\ \nonumber \\ \nonumber \\
  \begin{gathered}
    \begin{fmffile}{pic05}
      \begin{fmfgraph*}(40,40)
        \fmfset{arrow_len}{.3cm}
        \fmfleft{L1,L2}
        \fmfright{R1,R2}
        \fmf{dots_arrow}{L1,V}
        \fmf{dots_arrow}{L2,V}
        \fmf{dashes_arrow}{R1,V}
        \fmf{dashes_arrow}{R2,V}
        \fmflabel{$q_1,(\alpha)$}{R1}
        \fmflabel{$q_2,(\beta)$}{R2}
        \fmflabel{$p_1$}{L1}
        \fmflabel{$p_2$}{L2}
        \fmfdot{V}
      \end{fmfgraph*}
    \end{fmffile}
  \end{gathered}
  &=-\cfrac{i}{8\ce^2}\left[p_1\cdot p_2 +\cfrac73\,q_1\cdot q_2-\cfrac56\,(p_1+p_2)\cdot(q_1+q_2)\right]\,\delta^{(\alpha)(\beta)} ~. \nonumber
\end{align}

{\vspace{.5cm}}

The one-loop three-point function for $\psi$ evaluated by the effective action at the first order in $\ce^{-1}$ reads
\begin{align}
  \begin{gathered}
    \begin{fmffile}{pic06}
      \begin{fmfgraph*}(60,60)
        \fmfset{arrow_len}{.3cm}
        \fmftop{T}
        \fmfbottom{B1,B2}
        \fmf{dashes,left=.6,tension=.7}{A,B,C,A}
        \fmf{dots_arrow,label=$l$}{T,A}
        \fmf{dots_arrow,label=$p$}{B1,C}
        \fmf{dots_arrow,label=$q$}{B2,B}
        \fmfdot{A,B,C}
      \end{fmfgraph*}
    \end{fmffile}
  \end{gathered}
   =&\cfrac{\pi^2}{8\ce^3} \Big[ 4 p^2 q^2 l^2 C(l,p,0,0,0) - l^2(l^2+p^2+q^2-p\cdot q)B(l,0,0) \nonumber \\
    & -p^2(l^2+p^2+q^2-l\cdot q) B(p,0,0) -q^2(l^2+p^2+q^2-l\cdot p) B(q,0,0) \Big].
\end{align}
Here $B$ and $C$ are Passarino-Veltman integrals~\cite{Passarino:1978jh} defined in 
~\ref{the_appendix}. The expression has a non-vanishing divergent part proportional 
to the following operators:
\begin{align}
  \square \psi ~(\pd\psi)^2 ,\qquad \psi(\square\psi)^2 .
\end{align}
In full agreement with the logic presented above, the effective action~\eqref{the_effective_action} 
lacks these operators as they do not respect the nonlinear symmetry realization. Therefore the diagram 
dynamically breaks the original effective action symmetry. Another point that should be noted is the 
fact that these operators differ in dimension compared with the three-particle interaction operators 
in the effective action~\eqref{the_effective_action}. Consequently, they can be viewed as second order
terms in $\square/\ce^2$ expansion, i.e., as higher order terms in the momentum expansion.

However, three-points functions alone can hardly be considered as a sufficient evidence of anomalous 
behavior, as they all vanish on-shell due to kinematics. Because of this the correspondent anomalous 
contributions cannot be probed empirically. To proceed with the search for anomalies we are going to 
evaluate one-loop four-point functions given by the effective action at the first order in $\ce^{-1}$.

Four-point functions can have divergent parts which do not vanish on-shell. That can be directly 
shown via dimensional considerations. Namely, a one-loop four-particle function for $\psi$ can generate 
operator $\ce^{-N} (\pd\psi)^4 \psi^{N-4}$ that does not vanish on-shell. In the particular case of 
a four-point function for $\psi$ such a diagram reads (in terms of  the Mandelstam 
variables\footnote{$s=(p_1+p_2)^2, t = (p_1+q_1)^2, u = (p_1+q_2)^2$.}):

\begin{align}
  \left.
  \begin{gathered}
    \begin{fmffile}{pic07}
      \begin{fmfgraph*}(45,45)
        \fmfset{arrow_len}{.3cm}
        \fmfleft{L1,L2}
        \fmfright{R1,R2}
        \fmf{dots_arrow}{L1,DL}
        \fmf{dots_arrow}{L2,UL}
        \fmf{dots_arrow}{R1,DR}
        \fmf{dots_arrow}{R2,UR}
        \fmf{dashes,tension=.7}{DL,UL,UR,DR,DL}
        \fmfdot{DL,DR,UL,UR}
        \fmflabel{$p_1$}{L1}
        \fmflabel{$p_2$}{L2}
        \fmflabel{$q_1$}{R1}
        \fmflabel{$q_2$}{R2}
      \end{fmfgraph*}
    \end{fmffile}
  \end{gathered}
  \hspace{.5cm}
  \right|_\text{on-shell} \nonumber
\end{align}
\begin{align}
  =&\cfrac{i \pi^2}{256 \ce^4}\cfrac{1}{d-1} \left[ s(s\,d +2 t) B(p_1+p_2,0,0) + t(t\,d+2s) B(p_2-p_1,0,0)\right] .
\end{align}
The complete one-loop four-point $\psi$ function matrix element on-shell contains the following divergent part:
\begin{align}
  \left.  \begin{gathered}
    \begin{fmffile}{pic17}
      \begin{fmfgraph}(40,40)
        \fmfset{arrow_len}{.2cm}
        \fmfleft{L1,L2}
        \fmfright{R1,R2}
        \fmf{dots_arrow}{L1,DL}
        \fmf{dots_arrow}{L2,UL}
        \fmf{dots_arrow}{R1,DR}
        \fmf{dots_arrow}{R2,UR}
        \fmf{dashes,tension=.3}{DL,UL,UR,DR,DL}
        \fmfdot{DL,DR,UL,UR}
      \end{fmfgraph}
    \end{fmffile}
  \end{gathered}
  +
  \begin{gathered}
    \begin{fmffile}{pic18}
      \begin{fmfgraph}(40,40)
        \fmfset{arrow_len}{.2cm}
        \fmfleft{L1,L2}
        \fmfright{R1,R2}
        \fmf{dots_arrow}{L1,DL}
        \fmf{dots_arrow}{L2,UL}
        \fmf{dots_arrow}{R1,DR}
        \fmf{dots_arrow}{R2,UR}
        \fmf{dashes,tension=.4}{DL,UL,DR,UR,DL}
        \fmfdot{DL,DR,UL,UR}
      \end{fmfgraph}
    \end{fmffile}
  \end{gathered}
  +
  \begin{gathered}
    \begin{fmffile}{pic19}
      \begin{fmfgraph}(40,40)
        \fmfset{arrow_len}{.2cm}
        \fmfleft{L1,L2}
        \fmfright{R1,R2}
        \fmf{dots_arrow}{L1,DL}
        \fmf{dots_arrow}{L2,UL}
        \fmf{dots_arrow}{R1,DR}
        \fmf{dots_arrow}{R2,UR}
        \fmf{dashes,tension=.4}{UL,UR,DL,DR,UL}
        \fmfdot{DL,DR,UL,UR}
      \end{fmfgraph}
    \end{fmffile}
  \end{gathered}
  \right|_\text{on-shell} =-\cfrac{i\pi^2}{32\ce^4}~ (s^2+t^2 +u^2) \left[ \cfrac{1}{d-4} + \text{finite part} \right] .
\end{align}
Its divergent part is proportional to the following operators:
\begin{align}
  (\pd\psi)^4,\qquad \psi\pd_\mu\pd_\nu\psi\, \pd^\mu\psi\,\pd^\nu\psi ~.
\end{align}
They do not vanish on-shell, thus such an anomalous contribution can be probed empirically.

Nonetheless, these divergences can be eliminated via Ward identities. In full analogy with gauge 
theories, the model admits a symmetry which establishes certain relations between amplitudes, 
no matter the fact that the symmetry is realized in a nonlinear way.

To obtain Ward identities it is required to use the field transformation law. It is given by the 
following expression~\cite{Arbuzov:2019rcl}:
\begin{align}
  \begin{split}
    &\exp\left[\cfrac{i}{2}~ u^{(\mu)(\nu)}L_{(\mu)(\nu)} + i \theta D + i \theta^{(\mu)} K_{(\mu)}\right] \exp\left[i \phi D + i \sigma^{(\alpha)} K_{(\alpha)}\right] \\
    &= \exp\left[i \phi' D + i \sigma'{}^{(\alpha)} K_{(\alpha)}\right]\exp\left[\cfrac{i}{2}\, u'{}^{(\mu)(\nu)} L_{(\mu)(\nu)}\right].
  \end{split}
\end{align}
Here $u^{(\mu)(\nu)}$, $\theta^{(\mu)}$, and $\theta$ are the transformation parameters, while $\phi = \ce^{-1}\psi$ is a dimensionless field variable. The transformation law reads
\begin{align}\label{the_transformation_law_explicitly}
  \begin{split}
    &\phi' = \phi + \theta ~,\\
    &\sigma^\mu + \theta^\mu +\cfrac12\,\left(\sigma_\nu u^{\nu\mu} + \phi \theta^\mu - \sigma^\mu \theta\right) + \cdots = \sigma'{}^\mu-\cfrac12\, \sigma'{}_\nu u^{\nu\mu} + \cdots ~.
  \end{split}
\end{align}
In the expression for $\sigma$ we neglected terms containing higher orders of the transformation parameters.

In such a way the field $\psi$ on its own admits a shift symmetry. The Ward identities are generated by the infinitesimal action of the transformations on the Feynman integral. At the $\ce^{-1}$-level Ward identities for $\psi$ read
\begin{align}
  \int\mathcal{D}[\psi,\sigma^{(\alpha)}]\left[ \cfrac{1}{\ce} \left\{ (\pd\sigma)^2 + \cfrac12~\sigma\square\sigma\right\}\right] e^{i \Gamma} =0 .
\end{align}
This identity shows that all diagrams with at least one external $\psi$ line connected to the three-point function vanish due to the symmetry. Consequently, all diagrams discussed before vanish and anomalous contributions are eliminated.

The same argument does not hold for the four-particle function, as the Ward identities at the $\ce^{-2}$ level read
\begin{align}
  0=\int\mathcal{D}[\psi,\sigma^{(\alpha)}]\left[\pd^\mu \left\{\pd_\mu\psi\, \sigma^2 - \cfrac53\, \psi\,\sigma\pd_\mu\sigma\right\}+\cfrac75 \psi (\pd\sigma)^2-\cfrac53\, \pd\psi\, \sigma\pd\sigma  \right] \, e^{i\Gamma} ~.
\end{align}
Because of the gradient term the identity cannot exclude the corresponding interaction from the theory.

Consequently, one should search for one-loop anomalous amplitudes generated by the effective action \eqref{the_effective_action} at the second level in $\ce^{-1}$. Such amplitudes exist and they are given by the following:

\begin{align}
  \left.
  \begin{gathered}
    \begin{fmffile}{pic08}
      \begin{fmfgraph*}(50,50)
        \fmfset{arrow_len}{.3cm}
        \fmfleft{L1,L2}
        \fmfright{R1,R2}
        \fmf{dots_arrow}{L1,VL}
        \fmf{dots_arrow}{L2,VL}
        \fmf{dots_arrow}{R1,VR}
        \fmf{dots_arrow}{R2,VR}
        \fmf{dashes,left=1,tension=.3}{VL,VR,VL}
        \fmfdot{VL,VR}
        \fmflabel{$p_1$}{L1}
        \fmflabel{$p_2$}{L2}
        \fmflabel{$q_1$}{R1}
        \fmflabel{$q_2$}{R2}
      \end{fmfgraph*}
    \end{fmffile}
  \end{gathered}
  \hspace{.2cm}\right|_\text{on-shell}
  &=i \, \cfrac{25\pi^2}{64}\,\cfrac{1}{\ce^{4}} ~ s^2~ B(p_1+p_2,0,0) ~.
\end{align}
\begin{align}
  &
  \left.
  \begin{gathered}
    \begin{fmffile}{pic11}        
      \begin{fmfgraph}(35,35)
        \fmfset{arrow_len}{.2cm}
        \fmfleft{L1,L2}
        \fmfright{R1,R2}
        \fmf{dots_arrow}{L1,VL}
        \fmf{dots_arrow}{L2,VL}
        \fmf{dots_arrow}{R1,VR}
        \fmf{dots_arrow}{R2,VR}
        \fmf{dashes,left=1,tension=.3}{VL,VR,VL}
        \fmfdot{VL,VR}
      \end{fmfgraph}
    \end{fmffile}
  \end{gathered}
  +
  \begin{gathered}
    \begin{fmffile}{pic12}
      \begin{fmfgraph}(35,35)
        \fmfset{arrow_len}{.2cm}
        \fmfleft{L1,L2}
        \fmfright{R1,R2}
        \fmf{dots_arrow}{L1,VD}
        \fmf{dots_arrow}{L2,VU}
        \fmf{dots_arrow}{R1,VD}
        \fmf{dots_arrow}{R2,VU}
        \fmf{dashes,left=1,tension=.3}{VD,VU,VD}
        \fmfdot{VU,VD}
      \end{fmfgraph}
    \end{fmffile}
  \end{gathered}
  +
  \begin{gathered}
    \begin{fmffile}{pic13}
      \begin{fmfgraph}(35,35)
        \fmfset{arrow_len}{.2cm}
        \fmfleft{L1,L2}
        \fmfright{R1,R2}
        \fmf{dots_arrow,tension=1.5}{L1,VD}
        \fmf{dots_arrow,tension=1.5}{L2,VU}
        \fmf{phantom}{R1,VD}
        \fmf{phantom}{R2,VU}
        \fmf{dashes,left=.5,tension=.3}{VD,VU,VD}
        \fmfdot{VD,VU}
        \fmffreeze
        \fmf{dots_arrow,left=.5}{R2,VD}
        \fmf{dots_arrow,right=.5}{R1,VU}
      \end{fmfgraph}
    \end{fmffile}
  \end{gathered}
  \right|_\text{on-shell}=-i\,\cfrac{25\pi^2}{32}\,\cfrac{1}{\ce^4}~(s^2+t^2+u^2) \left[\cfrac{1}{d-4}+\text{finite part}\right] ~.
\end{align}
Divergences of these diagrams appear on-shell and can be probed empirically. At the same time they cannot be eliminated via Wards identities, so they lead to real effects. Finally, in full agreement with the symmetry reasoning operator $(\pd\psi)^4$ which is generated by the divergent part of the diagrams is missing in the original  effective action, as it does not respect the non-linear symmetry realization. 

It is required to make a comment on the renormalizability of the model. The fact that one-loop amplitudes generate operators missing in the original effective action~\eqref{the_effective_action_momentum_expansion} shows that the model cannot be renormalized by the standard technique. On the other hand it can only be considered as an effective model applicable in the low energy regime. The complete model which, indeed, should be renormalizable and free from anomalies must explicitly contain a mechanism of spontaneously breaking down the conformal symmetry and generating the energy scale $\ce$. The nonlinear symmetry realization technique used to generate the effective action~\eqref{the_effective_action} can only recover the dynamics of low energy modes. 
It is safe to assume that the divergent part of these diagrams can be renormalized when treated in the complete theory. It also should be highlighted that aforementioned diagrams alongside diagrams to be discussed further generate divergent local parts and finite non-local parts. Their role is discussed in the last section of the paper.

Finally, it is possible to recover the anomaly-induced effective action. To do this it is required to evaluate the rest one-loop four-particle diagrams:

\begin{align}
  \left.
  \begin{gathered}
    \begin{fmffile}{pic09}
      \begin{fmfgraph*}(50,50)
        \fmfset{arrow_len}{.3cm}
        \fmfleft{L1,L2}
        \fmfright{R1,R2}
        \fmf{dots_arrow}{L1,V}
        \fmf{dots_arrow}{L2,V}
        \fmf{dashes,tension=.5}{D,V,U}
        \fmf{dashes_arrow}{R1,D}
        \fmf{dashes_arrow}{R2,U}
        \fmf{dots,tension=.3,left=.3}{U,D}
        \fmflabel{$p_1$}{L1}
        \fmflabel{$p_2$}{L2}
        \fmflabel{$q_1$}{R1}
        \fmflabel{$q_2$}{R2}
        \fmfdot{V,U,D}
      \end{fmfgraph*}
    \end{fmffile}
  \end{gathered}
  \hspace{.2cm}\right|_\text{on-shell} &=-\cfrac{5 \pi^2}{64 \ce^4}~s^2 ~\left[\cfrac{1}{d-4} +\text{finite part}\right], \\ \nonumber \\
  \left.
  \begin{gathered}
    \begin{fmffile}{pic10}
      \begin{fmfgraph*}(45,45)
        \fmfset{arrow_len}{.3cm}
        \fmfleft{L1,L2}
        \fmfright{R1,R2}
        \fmf{dashes_arrow}{L1,L}
        \fmf{dashes_arrow}{L2,L}
        \fmf{dashes_arrow}{R1,R}
        \fmf{dashes_arrow}{R2,R}
        \fmf{phantom,tension=.6}{L,V,R}
        \fmfblob{25}{V}
        \fmfdot{L,R}
        \fmflabel{$p_1$}{L1}
        \fmflabel{$p_2$}{L2}
        \fmflabel{$q_1$}{R1}
        \fmflabel{$q_2$}{R2}
      \end{fmfgraph*}
    \end{fmffile}
  \end{gathered}
  \hspace{.2cm}\right|_\text{on-shell} &=
  \left.
  \begin{gathered}
    \begin{fmffile}{pic14}
      \begin{fmfgraph}(30,30)
        \fmfset{arrow_len}{.2cm}
        \fmfleft{L1,L2}
        \fmfright{R1,R2}
        \fmf{dashes_arrow}{L1,L}
        \fmf{dashes_arrow}{L2,L}
        \fmf{dashes_arrow}{R1,R}
        \fmf{dashes_arrow}{R2,R}
        \fmf{dots,left=1,tension=.3}{L,R,L}
        \fmfdot{L,R}  
      \end{fmfgraph}
    \end{fmffile}
  \end{gathered}
  +
  \begin{gathered}
    \begin{fmffile}{pic15}
      \begin{fmfgraph}(30,30)
        \fmfset{arrow_len}{.2cm}
        \fmfleft{L1,L2}
        \fmfright{R1,R2}
        \fmf{dashes_arrow}{L1,D}
        \fmf{dashes_arrow}{L2,U}
        \fmf{dashes_arrow}{R1,D}
        \fmf{dashes_arrow}{R2,U}
        \fmf{dots,left=1,tension=.3}{U,D,U}
        \fmfdot{U,D}
      \end{fmfgraph}
    \end{fmffile}
  \end{gathered}
  +
  \begin{gathered}
    \begin{fmffile}{pic16}
      \begin{fmfgraph}(30,30)
        \fmfset{arrow_len}{.2cm}
        \fmfleft{L1,L2}
        \fmfright{R1,R2}
        \fmf{dashes_arrow,tension=2}{L1,D}
        \fmf{dashes_arrow,tension=2}{L2,U}
        \fmf{phantom}{R1,D}
        \fmf{phantom}{R2,U}
        \fmf{dots,tension=.3,left=.5}{U,D,U}
        \fmffreeze
        \fmfdot{U,D}
        \fmf{dashes_arrow,right=.3}{R1,U}
        \fmf{dashes_arrow,left=.3}{R2,D}
      \end{fmfgraph}
    \end{fmffile}
  \end{gathered} 
  \right|_\text{on-shell} \nonumber\\
  &=-i \cfrac{25 \pi^2}{128 \ce^4}~(s^2+t^2+u^2) ~\left[\cfrac{1}{4-d}+\text{finite part}\right].
\end{align}

These expressions allow one to restore the anomaly-induced action:
\begin{align}\label{the_anomaly-induced_effective_action}
  \begin{split}
    \Gamma_\text{anomaly} = \int d^4 x \Bigg[ &-\cfrac12\,\psi\square\psi-\cfrac12\,\sigma\square\sigma +\cfrac{1}{4\ce^2}\Bigg\{\cfrac12\,\sigma^2(\pd\psi)^2+\cfrac76\,\psi^2(\pd\sigma)^2 -\cfrac53\,(\psi\pd\psi)(\sigma\pd\sigma)\Bigg\}\\
      & +\cfrac{1}{\ce^4} \Bigg\{ c_1 \, (\pd\psi)^4+c_{2,1} (\pd\psi)^2(\pd\sigma)^2+c_{2,2} \, (\pd\psi)^2 (\sigma\square\sigma) +c_{2,3} (\pd\sigma)^2 \, (\psi\square\psi) \\
      & +c_{3,1} (\pd\sigma)^4 +c_{3,2} (\pd\sigma)^2 (\sigma\square\sigma) \Bigg\}\Bigg].
  \end{split}
\end{align}

\section{Discussion and conclusion}

Results are analogous to the well-known case of the anomaly in the
$\sigma$-model~\cite{Adler:1969gk,Bell:1969ts}. In the low energy regime the original 
symmetry of the model is spontaneously broken. The effective action, nonetheless, respects 
the nonlinear realization, because of which some interactions do not appear in the model. 
Anomalous amplitudes have divergent parts proportional to operators missing in the effective action.
This feature is understood as a direct indication of a dynamical breaking of the effective action symmetry.

In full analogy with the given arguments one can conclude that an effective gravity action recovered via a nonlinear symmetry realization at the classical level should not be considered sufficient for a description of real low-energy gravitational processes. It should be used further to restore the corresponding anomaly-induced effective action which, in turn, should be implemented for calculations of physical observables. The studied model shows that an effective action and the corresponding anomaly-induced effective action can have not only different forms, but also different symmetries.

It should be noted that a similar anomalous behavior takes place in completely different gravity models. Namely, it appears also in Galileon models~\cite{Nicolis:2008in,Deffayet:2013lga}. In the flat spacetime Galileon models describe a scalar field with the so-called generalized Galilean symmetry. Because of the symmetry the theory admits second-order field equations and it is free from ghost states. However, at the one-loop level the theory generates higher derivative terms that break the original model 
symmetry~\cite{Brouzakis:2013lla,Heisenberg:2014raa}. Models of that class can be considered effective as the generalized Galilean symmetry can be induced by auxiliary dimensions~\cite{Nicolis:2008in,Deffayet:2013lga}. Similar phenomenology is preserved even in a curved spacetime. Within generalized Galileons (Horndeski models) one-loop effects generate both new interactions~\cite{Arbuzov:2017nhg} and higher-derivative terms~\cite{Latosh:2018xai}.

These considerations highlight the special role of symmetries in effective models. Quantum anomalies can hide the original symmetry of the model thereby complicating its empirical verification. Consequently, even if the conformal symmetry (or any other symmetry) is realized in nature in a nonlinear way, it still may not be probed easily due to anomalies.

Perhaps, the simplest reliable way to verify a theory with such anomalous features is given by non-local contributions of anomalous amplitudes. Such amplitudes generate finite non-local contributions alongside the divergent parts. For instance, within the model studied in this paper the one-loop four-point $\psi$ function generates the following non-local operator:

\begin{align}
  \begin{gathered}
    \begin{fmffile}{pic08}
      \begin{fmfgraph*}(50,50)
        \fmfset{arrow_len}{.3cm}
        \fmfleft{L1,L2}
        \fmfright{R1,R2}
        \fmf{dots_arrow}{L1,VL}
        \fmf{dots_arrow}{L2,VL}
        \fmf{dots_arrow}{R1,VR}
        \fmf{dots_arrow}{R2,VR}
        \fmf{dashes,left=1,tension=.3}{VL,VR,VL}
        \fmfdot{VL,VR}
        \fmflabel{$p_1$}{L1}
        \fmflabel{$p_2$}{L2}
        \fmflabel{$q_1$}{R1}
        \fmflabel{$q_2$}{R2}
      \end{fmfgraph*}
    \end{fmffile}
  \end{gathered}
  \to -i\,\cfrac{25\pi^2}{16\ce^4}~ p_1\cdot p_2 ~ q_1 \cdot q_2 \, \ln\left(-\cfrac{(p_1+p_2)^2}{\mu^2}\right) . 
\end{align}

This operator has a finite coupling and it is to affect real physical processes. Therefore the exact value of the coupling can be recovered from empirical data. These couplings are completely fixed by the original effective action~\eqref{the_effective_action} and they are not affected by the anomaly. It should also be noted that the role of non-local operators generated at the one-loop level is actively studied within effective gravity~\cite{Alexeyev:2017scq,Calmet:2015dpa,Donoghue:2014yha,Espriu:2005qn}.

Concluding, one can argue that anomalies of effective models with a nonlinear symmetry realization should be subjected to a more detailed consideration. The model studied it this paper alongside the other mentioned results shows that anomalies have non-trivial effects on the low-energy phenomenology.

\appendix

\section{Expressions for diagrams}\label{the_appendix}

The following definitions of Passarino-Veltman integrals are used~\cite{Passarino:1978jh,Mertig:1990an}:
\begin{align}
  B(p,m_1,m_2) \overset{\text{def}}{=}& \cfrac{(2\pi \mu)^{4-d}}{i\pi^2} \int d^dk ~\left[(k^2-m_1^2) \,((k-p)^2-m_2^2)\right]^{-1} \nonumber \\
  =& -\cfrac{2}{d-4} + \text{finite part} ~, \\
  C(p,p-q,q,m_1,m_2,m_3) \overset{\text{def}}{=}&\cfrac{(2\pi \mu)^{4-d}}{i\pi^2} \int d^d k ~\left[(k^2-m_1^2)\,((k-p)^2-m_2^2)\,((k-q)^2-m_3^2)\right]^{-1} .\nonumber
\end{align}

One-loop on-shell four-particle diagrams used in the paper are given by the following expressions:
\begin{align}
  \begin{gathered}
    \begin{fmffile}{psi4_1}
      \begin{fmfgraph*}(28,28)
        \fmfleft{L1,L2}
        \fmfright{R1,R2}
        \fmf{dots}{L1,L}
        \fmf{dots}{L2,L}
        \fmf{dots}{R1,R}
        \fmf{dots}{R2,R}
        \fmf{dashes,left=1,tension=.3}{L,R,L}
        \fmflabel{$p_1$}{L1}
        \fmflabel{$p_2$}{L2}
        \fmflabel{$q_1$}{R1}
        \fmflabel{$q_2$}{R2}
        \fmfdot{L,R}
      \end{fmfgraph*}
    \end{fmffile}
  \end{gathered}
  &=i \cfrac{25 \pi^2}{64 \ce^4}\, s^2\, B(p_1+p_2,0,0), &
  \begin{gathered}
    \begin{fmffile}{psi4_2}
      \begin{fmfgraph*}(28,28)
        \fmfleft{L1,L2}
        \fmfright{R1,R2}
        \fmf{dots}{L1,D}
        \fmf{dots}{L2,U}
        \fmf{dots}{R1,D}
        \fmf{dots}{R2,U}
        \fmf{dashes,tension=.3,left=1}{U,D,U}
        \fmfdot{U,D}
        \fmflabel{$p_1$}{L1}
        \fmflabel{$p_2$}{L2}
        \fmflabel{$q_1$}{R1}
        \fmflabel{$q_2$}{R2}
      \end{fmfgraph*}
    \end{fmffile}
  \end{gathered}
  &=i \,\cfrac{25\pi^2}{64\ce^4}\,\ t^2\, B(p_1+q_1,0,0), \\ \nonumber
\end{align}
\begin{align}
  \begin{gathered}
    \begin{fmffile}{psi4_3}
      \begin{fmfgraph*}(28,28)
        \fmfleft{L1,L2}
        \fmfright{R1,R2}
        \fmf{dots}{L1,D}
        \fmf{dots}{L2,U}
        \fmf{phantom}{R1,D}
        \fmf{phantom}{R2,U}
        \fmf{dashes,tension=.3,left=.5}{U,D,U}
        \fmffreeze
        \fmfdot{U,D}
        \fmf{dots,right=.4}{R1,U}
        \fmf{dots,left=.4}{R2,D}
        \fmflabel{$p_1$}{L1}
        \fmflabel{$p_2$}{L2}
        \fmflabel{$q_1$}{R1}
        \fmflabel{$q_2$}{R2}
      \end{fmfgraph*}
    \end{fmffile}
  \end{gathered}
  &= i\,\cfrac{25\pi^2}{64\ce^4}\,\ u^2 \, B(p_1+q_2,0,0), &
  \begin{gathered}
    \begin{fmffile}{sigma4_1}
      \begin{fmfgraph*}(28,28)
        \fmfleft{L1,L2}
        \fmfright{R1,R2}
        \fmf{dashes}{L1,L}
        \fmf{dashes}{L2,L}
        \fmf{dashes}{R1,R}
        \fmf{dashes}{R2,R}
        \fmf{dots,left=1,tension=.3}{L,R,L}
        \fmflabel{$p_1$}{L1}
        \fmflabel{$p_2$}{L2}
        \fmflabel{$q_1$}{R1}
        \fmflabel{$q_2$}{R2}
        \fmfdot{L,R}
      \end{fmfgraph*}
    \end{fmffile}
  \end{gathered}
  &=  i\, \cfrac{25\pi^2}{256\ce^4}\,s^2\, B(p_1+p_2,0,0), \nonumber \\ \nonumber
\end{align}
\begin{align}
  \begin{gathered}
    \begin{fmffile}{sigma4_2}
      \begin{fmfgraph*}(28,28)
        \fmfleft{L1,L2}
        \fmfright{R1,R2}
        \fmf{dashes}{L1,D}
        \fmf{dashes}{L2,U}
        \fmf{dashes}{R1,D}
        \fmf{dashes}{R2,U}
        \fmf{dots,left=1,tension=.3}{U,D,U}
        \fmflabel{$p_1$}{L1}
        \fmflabel{$p_2$}{L2}
        \fmflabel{$q_1$}{R1}
        \fmflabel{$q_2$}{R2}
        \fmfdot{U,D}
      \end{fmfgraph*}
    \end{fmffile}
  \end{gathered}
  &=i\, \cfrac{25\pi^2}{256\ce^4}\,t^2\, B(p_1+q_1,0,0),&
  \begin{gathered}
    \begin{fmffile}{sigma4_3}
      \begin{fmfgraph*}(28,28)
        \fmfleft{L1,L2}
        \fmfright{R1,R2}
        \fmf{dashes}{L1,D}
        \fmf{dashes}{L2,U}
        \fmf{phantom}{R1,D}
        \fmf{phantom}{R2,U}
        \fmf{dots,tension=.3,left=.5}{U,D,U}
        \fmffreeze
        \fmfdot{U,D}
        \fmf{dashes,right=.3}{R1,U}
        \fmf{dashes,left=.3}{R2,D}
        \fmflabel{$p_1$}{L1}
        \fmflabel{$p_2$}{L2}
        \fmflabel{$q_1$}{R1}
        \fmflabel{$q_2$}{R2}
      \end{fmfgraph*}
    \end{fmffile}
  \end{gathered}
  &=i\, \cfrac{25\pi^2}{256\ce^4}\,u^2\, B(p_1+q_2,0,0). \nonumber
\end{align}

\bibliographystyle{unsrturl}
\bibliography{Literaturverzeichnis}

\end{document}